\documentclass{article}
\usepackage{amsmath}
\usepackage[psamsfonts]{amssymb}
\usepackage{cmmib57}
\usepackage{diagrams}
\newcommand{\bPf}{\par\vspace*{-4pt}\indent{\sc Proof.}\enskip}
\newcommand{\ePf}{\medskip}
\def\QED{\hskip0.1em\hfill\null\ \null\nobreak\hfill\kern3pt\vbox{\hrule\hbox
    {\vrule\kern1pt\vbox{\kern1.7pt\hbox{$\scriptscriptstyle{QED}$}
     \kern0.2pt}\kern1pt\vrule}\hrule}}

\def\END{\hskip0.1em\hfill\null\ \null\nobreak\hfill\kern3pt\vbox{\hrule\hbox
    {\vrule\kern1pt\vbox{\kern1.7pt\hbox{$\,\,\,\vspace{5pt}$}
     \kern0.2pt}\kern1pt\vrule}\hrule}}
\newtheorem{theorem}{Theorem}
\newtheorem{lemma}{Lemma}
\newtheorem{corollary}{Corollary}
\newtheorem{proposition}{Proposition}
\newtheorem{remark}{Remark}
\newtheorem{definition}{Definition}
\newtheorem{example}{Example}
\newcommand{\bCd}{\bEq\begin{CD}}
\newcommand{\eCd}{\end{CD}\eEq}
\newcommand{\bcd}{\beq\begin{CD}}
\newcommand{\ecd}{\end{CD}\eeq}
\newcommand{\ben}{\begin{enumerate}}
\newcommand{\een}{\end{enumerate}}
\newcommand{\bEq}{\begin{eqnarray}}
\newcommand{\eEq}{\end{eqnarray}}
\newcommand{\beq}{\begin{eqnarray*}}
\newcommand{\eeq}{\end{eqnarray*}}
\newcommand{\bDf}{\begin{definition}\em}
\newcommand{\eDf}{\end{definition}}
\newcommand{\bLm}{\begin{lemma}}
\newcommand{\eLm}{\end{lemma}}
\newcommand{\bPr}{\begin{proposition}}
\newcommand{\ePr}{\end{proposition}}
\newcommand{\bTh}{\begin{theorem}}
\newcommand{\eTh}{\end{theorem}}
\newcommand{\bCr}{\begin{corollary}}
\newcommand{\eCr}{\end{corollary}}
\newcommand{\bRm}{\begin{remark}\em}
\newcommand{\eRm}{\end{remark}}
\newcommand{\bEx}{\begin{example}\em}
\newcommand{\eEx}{\end{example}}

\newcommand{\ie}{{\em i.e$.$} }
\newcommand{\eg}{{\em e.g$.$} }
\newcommand{\R}{I\!\!R}
\newcommand{\A}{\forall}

\newcommand{\mto}{\mapsto}

\newcommand{\der}{\partial}
\newcommand{\nab}{\nabla}

\DeclareMathOperator{\im}{im}
\DeclareMathOperator{\byd}{{\raisebox{.1ex}{:}{=}}}

\newcommand{\ucar}[1]{\underset{#1}{\times}}

\newcommand{\owed}[1]{\overset{#1}{\wedge}}

\newcommand{\balp}{\boldsymbol{\alp}}
\newcommand{\bbet}{\boldsymbol{\bet}}

\newcommand{\cA}{\mathcal{A}}

\newcommand{\cC}{\mathcal{C}}

\newcommand{\cE}{\mathcal{E}}

\newcommand{\cH}{\mathcal{H}}

\newcommand{\cJ}{\mathcal{J}}

\newcommand{\cL}{\mathcal{L}}

\newcommand{\cT}{\mathcal{T}}

\newcommand{\by}{\boldsymbol{y}}

\newcommand{\bF}{\boldsymbol{F}}
\newcommand{\bG}{\boldsymbol{G}}

\newcommand{\bP}{\boldsymbol{P}}

\newcommand{\bW}{\boldsymbol{W}}
\newcommand{\bX}{\boldsymbol{X}}
\newcommand{\bY}{\boldsymbol{Y}}

\newcommand{\sub}{\subset}

\newcommand{\wed}{\wedge}
\newcommand{\com}{\!\circ\!}
\newcommand{\ten}{\!\otimes\!}

\newcommand{\alp}{\alpha}
\newcommand{\bet}{\beta}
\newcommand{\gam}{\gamma}
\newcommand{\del}{\delta}
\newcommand{\eps}{\epsilon}
\newcommand{\zet}{\zeta}

\newcommand{\lam}{\lambda}
\newcommand{\sig}{\sigma}

\newcommand{\ome}{\omega}
\newcommand{\Gam}{\Gamma}

\newcommand{\Lam}{\Lambda}

\newcommand{\Ome}{\Omega}

\newcommand{\vartht}{\vartheta}

\newcommand{\For}{{\Lambda}}
\newcommand{\Con}{{\mathcal{C}}}
\newcommand{\Hor}{{\mathcal{H}}}
\newcommand{\Var}{{\mathcal{V}}}
\newcommand{\Thd}{{\Theta}}
\title{\large{{\bf Second variational derivative of gauge-natural 
invariant Lagrangians and conservation laws}\thanks{This work has 
been partially supported by
GNFM of INdAM, MIUR (PRIN 2003) and University of Torino.}}}
\author{{\normalsize M. Francaviglia, M.
Palese and E. Winterroth}
\\{\footnotesize Department of Mathematics,
University of Torino}
\\{\footnotesize via C. Alberto 10, 10123 Torino, Italy}\\
{\footnotesize e--mails:
{\sc francaviglia@dm.unito.it, palese@dm.unito.it, ekkehart@dm.unito.it}}}
\date{}
\begin{document}
\maketitle
\begin{abstract}
  We consider the second variational derivative of a given 
gauge-natural invariant Lagrangian taken with respect to 
(prolongations of) vertical parts of gauge-natural lifts of 
infinitesimal principal automorphisms. By requiring such a second 
variational derivative to vanish, {\em via} the Second Noether 
Theorem we find that a covariant strongly conserved current 
is
canonically
associated with the deformed Lagrangian obtained by contracting
Euler--Lagrange equations of the original Lagrangian with 
(prolongations of) vertical parts of gauge-natural lifts of 
infinitesimal principal automorphisms lying in the kernel of the 
generalized gauge-natural Jacobi morphism.
\medskip 


\end{abstract}

\section{Introduction}

After the works by Bergmann (see  \eg
\cite{Ber58} and references quoted therein)
the general problem has been tackled of coherently
defining the lifting of infinitesimal transformations of the basis 
manifold to bundles
of fields (namely bundles of tensors or tensor densities which 
could be obtained as
suitable representations of the action of infinitesimal space-time 
transformations on frame
bundles of a given order \cite{PaTe77}). Such theories were also 
called geometric or
{\em natural} \cite{Tra67}. A physically important generalization of natural 
theories to gauge field theories passed
through
the concept of jet prolongation of a principal bundle and the
introduction of a very important geometric construction, namely the 
{\em gauge-natural} bundle
functor \cite{Ec81,KMS93}.

We recall that within the above mentioned program generalized Bianchi 
identities for geometric field
theories were introduced by Bergmann to get (after an  integration by 
parts procedure) a consistent
equation involving divergences within the first variation formula. It was
also stressed that in the general theory of Relativity those identities
coincide with the  contracted Bianchi identities for the curvature tensor of
the pseudo-Riemannian metric. Our present aim is to suitably generalize this setting to the
gauge-natural framework.

Our general setting is the calculus of variations on finite order fibered
bundles which will be assumed to be
{\em gauge-natural bundles} (\ie jet prolongations of fiber bundles associated
to some gauge-natural prolongation of a principal bundle $\bP$
\cite{Ec81,KMS93}).
Such geometric structures have been widely recognized to suitably describe
so-called gauge-natural field theories, \ie physical theories in which
right-invariant infinitesimal automorphisms of the structure bundle $\bP$
uniquely define the transformation laws of the fields themselves (see 
{\em e.g.}
\cite{FaFr03} and references quoted therein).
We
shall in particular consider {\em finite order variational sequences} 
\cite{Kru90}
on gauge-natural bundles.
We represent the second variational 
derivative \cite{FrPa00,FrPa01,FPV05} of a given gauge-natural 
invariant Lagrangian taken with respect to (prolongations of) 
vertical parts of gauge-natural lifts of infinitesimal principal 
automorphisms.
We define the generalized gauge-natural Jacobi morphism associated 
with a given gauge-natural invariant Lagrangian
by taking as  {\em variation vector fields}
the Lie derivatives
of sections of the gauge-natural bundle with respect to gauge-natural 
lifts of infinitesimal
automorphisms of the underlying principal bundle $\bP$. Such variation vector fields are, in
particular, generalized  symmetries \cite{cacca,GMS03}.
Recall that, as a consequence of the Second Noether Theorem 
\cite{Noe18}, within such a picture it is possible to  relate the 
generalized Bianchi
morphism to the second variational derivative of the Lagrangian and 
then to the associated generalized gauge-natural Jacobi morphism; 
details will appear in
\cite{PaWi03,PaWi04}. 

Recall that, in the
case of geodesics in a Riemannian manifold, vector fields which
make the second variation to vanish identically modulo boundary terms are
called {\em Jacobi fields} and they are solutions of a second order
differential equation known as {\em Jacobi equation} (of geodesics).
The notion of Jacobi equation as an outcome of the second variation is in
fact fairly more general: formulae for the second
variation of a Lagrangian functional in higher order field theory and 
generalized Jacobi equations along critical sections have been
already considered (see, \eg the results of \cite{Vari,CFT96,GoSt73} 
and classical references quoted therein).

In particular, by requiring the gauge-natural second variational 
derivative to vanish, we find that a covariant Noether strongly 
conserved current is
canonically
associated with the generalized {\em deformed} Lagrangian $\ome$ 
obtained by contracting
Euler--Lagrange equations of the original Lagrangian $\lam$ with 
(prolongations of) vertical parts of gauge-natural lifts of 
infinitesimal principal automorphisms which are in the kernel of the 
generalized gauge-natural Jacobi morphism.  

Concluding remarks are dedicated to the comparison of our results with the ones reported in
the recent paper \cite{GMS03},  where conservation laws associated with generalized
Lagrangian  symmetries, and in particular with divergence symmetries of a 
Lagrangian
are presented, {\em via} the so-called {\em characteristic equation}, whithin the framework
of infinite order variational bicomplexes.  

\section{Finite order jets of gauge-natural bundles and variational sequences}

We recall some basic facts about jet spaces, sheaves of forms on the 
$s$--th order
jet space (standard references are \eg \cite{KMS93,Kru90,Kru93,Sau89,Vit98,VitAPPL}) 
and we mainly set the notation.

Our framework is a fibered manifold $\pi : \bY \to \bX$,
with $\dim \bX = n$ and $\dim \bY = n+m$.
For $s \geq q \geq 0$ integers we are concerned with the $s$--jet 
space $J_s\bY$ of $s$--jet prolongations of (local) sections
of $\pi$; in particular, we set $J_0\bY \equiv \bY$. We recall the 
natural fiberings
$\pi^s_q: J_s\bY \to J_q\bY$, $s \geq q$, $\pi^s: J_s\bY \to \bX$, and,
among these, the {\em affine\/} fiberings $\pi^{s}_{s-1}$.
We denote by $V\bY$ the vector subbundle of the tangent
bundle $T\bY$ of vectors on $\bY$ which are vertical with respect
to the fibering $\pi$.
Greek
indices $\sig ,\mu ,\dots$ run from $1$ to $n$ and they label basis
coordinates, while
Latin indices $i,j,\dots$ run from $1$ to $m$ and label fibre coordinates,
unless otherwise specified.
We denote multi--indices of dimension $n$ by boldface Greek letters such as
$\balp = (\alp_1, \dots, \alp_n)$, with $0 \leq \alp_\mu$,
$\mu=1,\ldots,n$; by an abuse
of notation, we denote by $\sig$ the multi--index such that
$\alp_{\mu}=0$, if $\mu\neq \sig$, $\alp_{\mu}= 1$, if
$\mu=\sig$.
We also set $|\balp| \byd \alp_{1} + \dots + \alp_{n}$ and $\balp ! \byd
\alp_{1}! \dots \alp_{n}!$.
The charts induced on $J_s\bY$ are denoted by $(x^\sig,y^i_{\balp})$, with $0
\leq |\balp| \leq s$; in particular, we set $y^i_{\bf{0}}
\equiv y^i$. The local vector fields and forms of $J_s\bY$ induced by
the above coordinates are denoted by $(\der^{\balp}_i)$ and $(d^i_{\balp})$,
respectively.

For $s\geq 1$, we consider the natural complementary fibered
morphisms over $J_s\bY \to J_{s-1}\bY$ (see \cite{Kru90,Kru93,Vit98}):
\beq
\mathcal{D} : J_s\bY \ucar{\bX} T\bX \to TJ_{s-1}\bY \,,
\qquad
\vartht : J_{s}\bY \ucar{J_{s-1}\bY} TJ_{s-1}\bY \to VJ_{s-1}\bY \,,
\eeq
with coordinate expressions, for $0 \leq |\balp| \leq s-1$, given by
\beq
\mathcal{D} = d^\lam\ten {\mathcal{D}}_\lam = d^\lam\ten
(\der_\lam + y^j_{\balp+\lam}\der_j^{\balp}) \,,\quad
\vartht = \vartht^j_{\balp}\ten\der_j^{\balp} =
(d^j_{\balp}-y^j_{{\balp}+\lam}d^\lam)
\ten\der_j^{\balp} \,.
\eeq

The morphisms above induce the following natural splitting (and its dual):
\bEq
\label{jet connection}
J_{s}\bY\ucar{J_{s-1}\bY}T^*J_{s-1}\bY =\left(
J_s\bY\ucar{J_{s-1}\bY}T^*\bX\right) \oplus\cC^{*}_{s-1}[\bY]\,,
\eEq
where $\cC^{*}_{s-1}[\bY] \byd \im \vartht_s^*$ and
$\vartht_s^* : J_s\bY \ucar{J_{s-1}\bY} V^*J_{s-1}\bY \to
J_s\bY \ucar{J_{s-1}\bY} T^*J_{s-1}\bY \,$.

If $f: J_{s}\bY \to \R$ is a function, then we set
$D_{\sig}f$ $\byd \mathcal{D}_{\sig} f$,
$D_{\balp+\sig}f$ $\byd D_{\sig} D_{\balp}f$, where $D_{\sig}$ is
the standard {\em formal derivative}.
Given a vector field $u: J_{s}\bY \to TJ_{s}\bY$, the splitting
\eqref{jet connection} yields $u \, \com \, \pi^{s+1}_{s} = u_{H} + u_{V}$
where, if $u = u^{\gam}\der_{\gam} + u^i_{\balp}\der^{\balp}_i$, then we
have $u_{H} = u^{\gam}D_{\gam}$ and
$u_{V} = (u^i_{\balp} - y^i_{\balp + \gam}u^{\gam})
\der^{\balp}_{i}$. We shall call $u_{H}$ and $u_{V}$ the
horizontal and the vertical part of $u$, respectively.

The splitting
\eqref{jet connection} induces also a decomposition of the
exterior differential on $\bY$,
$(\pi^{s}_{s-1})^*\com \,d = d_H + d_V$, where $d_H$ and $d_V$
are defined to be the {\em horizontal\/} and {\em vertical differential\/}.
The action of $d_H$ and $d_V$ on functions and $1$--forms
on $J_s\bY$ uniquely characterizes $d_H$ and $d_V$ (see, {\em e.g.},
\cite{Kru90,Kru93,Sau89,Vit98} for more details).
A {\em projectable vector field\/} on $\bY$ is defined to be a pair
$(\Xi,\xi)$, where $\Xi:\bY \to T\bY$ and $\xi: \bX \to T\bX$
are vector fields and $\Xi$ is a fibered morphism over $\xi$.
If there is no danger of confusion, we will denote simply by $\Xi$ a
projectable vector field $(\Xi,\xi)$.
A projectable vector field $(\Xi,\xi)$
can be prolonged by the flow functor to a projectable vector field
$(j_{s}\Xi, \xi)$, the coordinate expression of which can be found \eg in
\cite{Kru90,Sau89,Vit98};
in particular, if locally $\Xi= 
\xi^{\sig}(x^{\mu})\der_{\sig}+\xi^{i}(x^{\mu},y^{j})\der_{i}$, we 
have the following expressions
$(j_{s}\Xi)_{H} = \xi^{\sig} \, D_{\sig}$,
$(j_{s}\Xi)_{V} = D_{\balp}(\Xi_{V})^i \, \der_i^{\balp}$,
with $(\Xi_{V})^i = \xi^i - \, y^i_{\sig}\xi^{\sig}$, for the
horizontal and the vertical part of $j_{s}\Xi$, respectively.
 From now on, by an abuse of notation, we will write simply $j_{s}\Xi_{H}$ and
$j_{s}\Xi_{V}$.
In particular, we stress that $j_{s}\Xi_{V}$ can be seen as a fibered 
morphism: $j_{s}\Xi_{V}: J_{s+1}\bY\ucar{J_{s}\bY}J_{s}\bY\to 
J_{s+1}\bY\ucar{J_{s}\bY}J_{s}V\bY$.

\subsection{Gauge-natural bundles}

Let $\bP\to\bX$ be a principal bundle with structure group $\bG$.
Let $r\leq k$ be integers and $\bW^{(r,k)}\bP$ $\byd$ 
$J_{r}\bP\ucar{\bX}L_{k}(\bX)$,
where $L_{k}(\bX)$ is the bundle of $k$--frames
in $\bX$ \cite{Ec81,KMS93}, $\bW^{(r,k)}\bG \byd J_{r}\bG \odot GL_{k}(n)$
the semidirect product with respect to the action of $GL_{k}(n)$
on $J_{r}\bG$ given by the
jet composition and $GL_{k}(n)$ is the group of $k$--frames
in $\R^{n}$. Here we denote by $J_{r}\bG$ the space of 
$(r,n)$-velocities on $\bG$.
The bundle $\bW^{(r,k)}\bP$ is a principal bundle over $\bX$ with 
structure group
$\bW^{(r,k)}\bG$.
Let $\bF$ be any manifold and $\zet:\bW^{(r,k)}\bG \ucar{}\bF\to\bF$ be
a left action of $\bW^{(r,k)}\bG$ on $\bF$. There is a naturally defined
right action of $\bW^{(r,k)}\bG$ on $\bW^{(r,k)}\bP \times \bF$ so that
  we can associate in a standard way
to $\bW^{(r,k)}\bP$ the bundle, on the given basis $\bX$,
$\bY_{\zet} \byd \bW^{(r,k)}\bP\times_{\zet}\bF$.

\bDf
We say $(\bY_{\zet},\bX,\pi_{\zet};\bF,\bG)$ to be the
{\em gauge-natural bundle} of order
$(r,k)$ associated to the principal bundle $\bW^{(r,k)}\bP$
by means of the left action $\zet$ of the group
$\bW^{(r,k)}\bG$ on the manifold $\bF$ \cite{Ec81,KMS93}.
\END\eDf

\bRm
A principal automorphism $\Phi$ of $\bW^{(r,k)}\bP$ induces an
automorphism of the gauge-natural bundle by:
\bEq
\Phi_{\zet}:\bY_{\zet}\to\bY_{\zet}: [(j^{x}_{r}\gam,j^{0}_{k}t),
\hat{f}]_{\zet}\mto [\Phi(j^{x}_{r}\gam,j^{0}_{k}t),
\hat{f}]_{\zet}\,,
\eEq
where $\hat{f}\in \bF$ and $[\cdot, \cdot]_{\zet}$ is the equivalence class
induced by the action $\zet$.
\END\eRm
\bDf
We also define the {\em vector} bundle  over $\bX$ of right invariant
infinitesimal automorphisms of $\bW^{(r,k)}\bP$ by setting
$\cA^{(r,k)} \byd T\bW^{(r,k)}\bP/\bW^{(r,k)}\bG$ ($r\leq k$).
\END\eDf

Denote by $\cT_{\bX}$ and $\cA^{(r,k)}$ the sheaf of
vector fields on $\bX$ and the sheaf of right invariant vector fields
on $\bW^{(r,k)}\bP$, respectively. A functorial mapping 
$\mathfrak{G}$ is defined
which lifts any right--invariant local automorphism $(\Phi,\phi)$ of the
principal bundle $W^{(r,k)}\bP$ into a unique local automorphism
$(\Phi_{\zet},\phi)$ of the associated bundle $\bY_{\zet}$.
Its infinitesimal version associates to any $\bar{\Xi} \in \cA^{(r,k)}$,
projectable over $\xi \in \cT_{\bX}$, a unique {\em projectable} vector field
$\hat{\Xi} \byd \mathfrak{G}(\bar{\Xi})$ (called the gauge-natural lift) on 
$\bY_{\zet}$ in the
following way:
\bEq
\mathfrak{G} : \bY_{\zet} \ucar{\bX} \cA^{(r,k)} \to T\bY_{\zet} \,:
(\by,\bar{\Xi}) \mto \hat{\Xi} (\by) \,,
\eEq
where, for any $\by \in \bY_{\zet}$, one sets: $\hat{\Xi}(\by)=
\frac{d}{dt} \left((\Phi_{\zet \,t})(\by)\right)_{t=0}$,
and $\Phi_{\zet \,t}$ denotes the (local) flow corresponding to the
gauge-natural lift of $\Phi_{t}$.

This mapping fulfils the following properties (see \cite{KMS93}):
\begin{enumerate}
\item $\mathfrak{G}$ is linear over $id_{\bY_{\zet}}$;
\item we have $T\pi_{\zet}\circ\mathfrak{G} = id_{T\bX}\circ
\bar{\pi}^{(r,k)}$,
where $\bar{\pi}^{(r,k)}$ is the natural projection
$\bY_{\zet}\ucar{\bX}
\cA^{(r,k)} \to T\bX$;
\item for any pair $(\bar{\Lam},\bar{\Xi})$ $\in$
$\cA^{(r,k)}$ we have
$\mathfrak{G}([\bar{\Lam},\bar{\Xi}]) = [\mathfrak{G}(\bar{\Lam}), 
\mathfrak{G}(\bar{\Xi})]$ for commutators.
\end{enumerate}

\subsection{Lie derivative of gauge-natural sections}

\bDf
Let $\gam$ be a (local) section of $\bY_{\zet}$, $\bar{\Xi}$
$\in \cA^{(r,k)}$ and $\hat\Xi$ its gauge-natural lift.
Following \cite{KMS93} we
define the {\em
generalized Lie derivative} of $\gam$ along the vector field
$\hat{\Xi}$ to be the (local) section $\pounds_{\bar{\Xi}} \gam : \bX 
\to V\bY_{\zet}$,
given by
\beq
\pounds_{\bar{\Xi}} \gam = T\gam \circ \xi - \hat{\Xi} \circ 
\gam \,.
\eeq\END
\eDf

\bRm\label{lie}
The Lie derivative operator acting on sections of gauge-natural
bundles satisfies the following
properties:
\begin{enumerate}\label{lie properties}
\item for any vector field $\bar{\Xi} \in \cA^{(r,k)}$, the
mapping $\gam \mto \pounds_{\bar{\Xi}}\gam$
is a first--order quasilinear differential operator;
\item for any local section $\gam$ of $\bY_{\zet}$, the mapping
$\bar{\Xi} \mto \pounds_{\bar{\Xi}}\gam$
is a linear differential operator;
\item we can regard $\pounds_{\bar{\Xi}}: J_{1}\bY_{\zet} \to V\bY_{\zet}$
as a morphism over the
basis $\bX$. By using the canonical
isomorphisms $VJ_{s}\bY_{\zet}\simeq J_{s}V\bY_{\zet}$ for all $s$, we have
$\pounds_{\bar{\Xi}}[j_{s}\gam] = j_{s} [\pounds_{\bar{\Xi}} \gam]$,
for any (local) section $\gam$ of $\bY_{\zet}$ and for any (local)
vector field $\bar{\Xi}\in \cA^{(r,k)}$. Furthermore, the fundamental 
relation holds true:
\beq
{\hat{\Xi}}_{V}\byd\mathfrak{G}(\bar{\Xi})_V =- 
\pounds_{\bar{\Xi}}\, .
\eeq \END
\end{enumerate}
\eRm

\subsection{Variational sequences}

For the sake of simplifying notation, we will sometimes omit the 
subscript $\zet$, so
that all our considerations shall refer to $\bY$ as a gauge-natural
bundle as defined above.

We shall be here concerned with some distinguished sheaves of forms on jet
spaces \cite{Kru90,Sau89,Vit98}. We shall in particular follow 
notation given in \cite{Vit98} to which the
reader is referred for details.
  For $s \geq 0$, we consider the standard sheaves $\For^{p}_{s}$
of $p$--forms on $J_s\bY$.
  For $0 \leq q \leq s $, we consider the sheaves $\Hor^{p}_{(s,q)}$ and
$\Hor^{p}_{s}$ of {\em horizontal forms} with respect to the 
projections $\pi^s_q$ and $\pi^s_0$, respectively.
  For $0 \leq q < s$, we consider the subsheaves $\Con^{p}_{(s,q)}
\sub \Hor^{p}_{(s,q)}$ and $\Con^{p}{_s} \sub
\Con^{p}_{(s+1,s)}$ of {\em contact forms}, \ie horizontal forms 
valued into $\cC^{*}_{s}[\bY]$(they have the property of vanishing 
along any section of the gauge-natural bundle).

According to \cite{Kru90,Vit98} the fibered splitting
\eqref{jet connection} yields the {\em sheaf splitting}
$\Hor^{p}_{(s+1,s)}$ $=$ $\bigoplus_{t=0}^p$
$\Con^{p-t}_{(s+1,s)}$ $\wed\Hor^{t}_{s+1}$, which restricts to the inclusion
$\For^{p}_s$ $\sub$ $\bigoplus_{t=0}^{p}$
$\Con^{p-t}{_s}\wed\Hor^{t,}{_{s+1}^{h}}$,
where $\Hor^{p,}{_{s+1}^{h}}$ $\byd$ $h(\For^{p}_s)$ for $0 < p\leq 
n$ and the surjective map
$h$ is defined to be the restriction to $\For^{p}_{s}$ of the projection of
the above splitting onto the non--trivial summand with the highest
value of $t$. By an abuse of notation, let us denote by $d\ker h$ the sheaf
generated by the presheaf $d\ker h$ in the standard way.
We set $\Thd^{*}_{s}$ $\byd$ $\ker h$ $+$
$d\ker h$.

In \cite{Kru90}, to which the reader is referred for details,
it was proved that the following sequence is an exact resolution of 
the constant sheaf $\R_{\bY}$ over $\bY$:
\beq
\diagramstyle[size=1.3em]
\begin{diagram}
0 & \rTo & \R_{Y} & \rTo & \For^{0}_s & \rTo^{\cE_{0}} &
\For^{1}_s/\Thd^{1}_s & \rTo^{\cE_{1}} & \For^{2}_s/\Thd^{2}_s & 
\rTo^{\cE_{2}} &
\dots & \rTo^{\cE_{I-1}} & \For^{I}_s/\Thd^{I}_s & \rTo^{\cE_{I}} &
\For^{I+1}_s & \rTo^{d} & 0
\end{diagram}
\eeq

\bDf
The above sequence, where the highest integer $I$ depends on the dimension
of the fibers of $J_{s}\bY \to \bX$ (see, in particular, 
\cite{Kru90}), is said to be the $s$--th order
{\em variational sequence\/} associated with the fibered manifold
$\bY\to\bX$.
\END
\eDf

For practical purposes we
shall limit ourselves to consider the truncated variational sequence:
\beq
\diagramstyle[size=1.3em]
\begin{diagram}
0 &\rTo & \R_{Y} &\rTo & \Var^{0}_s & \rTo^{\cE_0} &
\Var^{1}_{s} & \rTo^{\cE_{1}} & \dots  & \rTo^{\cE_{n}} &
\Var^{n+1}_{s}  & \rTo^{\cE_{n+1}} & \cE_{n+1}(\Var^{n+1}_{s})
& \rTo^{\cE_{n+2}} &
0 \,,
\end{diagram}
\eeq
where, following \cite{Vit98}, the sheaves $\Var^{p}_{s}\byd
\Con^{p-n}_{s}\wed\Hor^{n,}{_{s+1}^h}/h(d\ker h)$, with $0\leq p\leq n+2$, are
suitable representations of the corresponding quotient
sheaves in the variational sequence by means of sheaves of sections of tensor
bundles. Representations of Krupka's long  variational sequence by 
means of differential forms have been provided \eg in 
\cite{KRBMusilROMP,Sedenkova,VitAPPL}.

  We recall now some intrinsic decomposition involved with the first 
and the second variation formulae. 
\begin{itemize}
\item Let 
$\alp\in\Con^{1}_s\wed\Hor^{n,}{_{s+1}^h}
\sub \Var^{n+1}_{s+1}$. Then there is a unique pair of
sheaf morphisms (\cite{Kol83,KoVi03,Vit98})
\bEq\label{first variation}
E_{\alp} \in \Con^{1}_{(2s,0)}\wed\Hor^{n,}{_{2s+1}^{h}} \,,
\qquad
F_{\alp} \in \Con^{1}_{(2s,s)} \wed \Hor^{n,}{_{2s+1}^h} \,,
\eEq
such that
$(\pi^{2s+1}_{s+1})^*\alp=E_{\alp}-F_{\alp}$
and $F_\alp$ is {\em locally} of the form $F_{\alp} = d_{H}p_{\alp}$, 
with $p_{\alp}
\in \Con^{1}_{(2s-1,s-1)}\wed\Hor^{n-1}{_{2s}}$.

\item Let 
$\eta\in\Con^{1}_{s}\wed\Con^{1}_{(s,0)}\wed\Hor^{n,}{_{s+1}^{h}}\sub
\Var^{n+2}_{s+1}$;
then there is a unique morphism
$$
K_{\eta} \in 
\Con^{1}_{(2s,s)}\otimes\Con^{1}_{(2s,0)}\wed\Hor^{n,}{_{2s+1}^{h}}
$$
such that, for all $\Xi:\bY\to V\bY$,
$
E_{{j_{s}\Xi}\rfloor \eta} = C^{1}_{1} (j_{2s}\Xi\ten K_{\eta})$,
where $C^1_1$ stands for tensor
contraction on the first factor and $\rfloor$ denotes inner product 
(see \cite{KoVi03,Vit98}).
Furthermore, there is a unique pair of sheaf morphisms
\bEq\label{second}
H_{\eta} \in
\Con^{1}_{(2s,s)}\wed\Con^{1}_{(2s,0)}\wed\Hor^{n,}{_{2s+1}^{h}} \,,
\quad
G_{\eta} \in \Con^{2}_{(2s,s)}\wed\Hor^{n,}{_{2s+1}^{h}} \,,
\eEq
such that
${(\pi^{2s+1}_{s+1})}^*\eta = H_{\eta} - G_{\eta}$ and $H_{\eta}
= \frac{1}{2} \, A(K_{\eta})$,
where $A$ stands for antisymmetrisation.
Moreover, $G_{\eta}$ is {\em locally} of the type $G_{\eta} = d_H q_{\eta}$,
where
$q_{\eta} \in \Con^{2}_{(2s-1,s-1)}\wed\Hor^{n-1}{_{2s}}$; hence
$[\eta]=[H_{\eta}]$ (see \cite{KoVi03,Vit98}).
\end{itemize}

\bRm
A section $\lam\in\Var^{n}_s$ is just a Lagrangian of order
$(s+1)$ of
the standard literature, while
$\cE_{n}(\lam) \in \Var^{n+1}_{s}$ coincides with the standard higher
order Euler--Lagrange morphism associated with $\lam$. The kernel of 
the morphism $H$ coincides with standard Helmholtz  conditions of 
local variationality.
\END
\eRm

\bEx\label{Ex1}
Let $\lam\in\Var^{n}_s$.
It is known (see {\em e.g.} \cite{Kol83}) that
\beq
& d_{V}\lam = (d_{V}\lam)^{\balp}_{i}\vartht^{i}_{\balp}\wed \ome \,,
\,
E_{d_{V}\lam}
= \cE_n (\lam)_{i}\vartht^{i}\wed \ome \,,
\,
p_{d_{V}\lam}
= p(\lam)^{\balp\mu}_{i}\vartht^{i}_{\balp}\wed \ome_{\mu}\,,
\\
& p(\lam)^{\bbet\mu}_{i}
= (d_{V}\lam)^{\balp}_{i} \qquad \bbet +\mu= \balp, |\balp|= s  \,,
\\
& p(\lam)^{\bbet\mu}_{i}
= (d_{V}\lam)^{\balp}_{i} - D_{\nu}p(\lam)^{\balp \nu}_{i}
\qquad \bbet+\mu=\balp, |\balp|= s-1  \,,
\\
& \cE_n (\lam)^{\balp}_{i}
= (d_{V}\lam)^{\balp}_{i}- D_{\nu}p(\lam)^{\balp\nu}_{i}\qquad |\balp|=0 \,.
\eeq
Furthermore,
$\cE_n (\lam)_{i}=\sum_{|\balp|\leq s}(-1)^{|\balp|}
D_{\balp}(d_{V}\lam)^{\balp}_{i}$.
\END 
\eEx

\section{Variations and generalized Jacobi morphisms}

We will represent generalized {\em gauge-natural} Jacobi morphisms in
variational sequences and establish their relation with the second 
variational derivative of a
generalized gauge-natural invariant Lagrangian.
We consider {\em formal variations} of a fibered morphism as
{\em multiparameter deformations} and relate
  the second variational derivative  of
the Lagrangian $\lam$ to the
variational Lie derivative of the associated Euler--Lagrange morphism 
and to the generalized
Bianchi morphism; see \cite{PaWi03} for details.

\bDf\label{var}
Let $\pi: \bY\to \bX$ be any bundle and let  $\alp: J_{s}\bY\to
\owed{p}T^*J_{s}\bY$ and $L_{j_{s}\Xi_{k}}$ be the Lie derivative
operator acting on differential fibered morphism.
Let
  $\Xi_{k}$, $1\leq k\leq i$, be variation
vector fields on $\bY$ in the sense of 
\cite{FrPa00,FrPa01,FPV05,PaWi03}. In particular in the sequel we 
will assume that they are vertical parts of projectable vector 
fields. Inspired by the classical approach \cite{GoSt73} we define 
the
$i$--th formal variation of the morphism $\alp$ to be the operator:
$\del^{i} \alp = L_{j_{s}\Xi_{1}} \ldots L_{j_{s}\Xi_{i}} \alp$.
\eDf

\bRm
Let now the morphism $\alp$ be a representative of an equivalence 
class in the variational sequence, \ie $\alp\in (\Var^{n}_{s})_{\bY}$.

Since the Lie derivative operator $L_{j_{s}\Xi}$ with respect to a projectable 
vector field $j_{s}\Xi$ preserves the contact 
structure \cite{FPV98a,Kru93}, by an abuse of notation we can 
write:
$\del^{i}[\alp]\byd [\del^{i}\alp]$  $=$ $[L_{\Xi_{i}} \ldots
L_{\Xi_{1}}\alp]$ $=$ $\cL_{\Xi_{i}} \ldots
\cL_{\Xi_{1}}[\alp]$, where $\cL_{\Xi_{i}}$ is the {\em variational 
Lie derivative} acting on quotient variational morphisms, an operator 
defined and conveniently represented in \cite{FPV98a}; see also 
\cite{Kru93}.\END
\eRm

\bDf
We call the operator $\del^{i}$ the {\em $i$--th  variational
derivative}. \END
\eDf

The following
characterization of the second variational derivative of a 
generalized Lagrangian
in the variational sequence holds true \cite{PaWi03}.

\bPr\label{x}
Let $\lam\in (\Var^{n}_{s})_{\bY}$ and let $\Xi$ be a variation vector
field; then we have
\bEq
\del^{2}\lam = [\cE_{n}(j_{2s}\Xi \rfloor h\del\lam)
+C^{1}_{1} (j_{2s}\Xi \ten K_{hd\del\lam})] \,.
\eEq
\ePr

\subsection{Generalized {\em gauge-natural} Jacobi morphisms}

Let now specify $\bY$ to be a gauge-natural bundle and let 
$\hat{\Xi}_V\equiv \mathfrak{G}(\bar{\Xi})_V$ be a variation vector 
field. Let us consider
$j_{s}{\bar{\Xi}}_{V}$,
\ie the vertical part according to the splitting \eqref{jet 
connection}. The set of all sections of the
vector bundle
$\cA^{(r+s,k+s)}$ of this kind defines a vector subbundle of 
$J_{s}\cA^{(r,k)}$, which (since we are speaking  about vertical parts with respect to the 
splitting \eqref{jet
connection}) by a slight abuse of notation will be denoted by
$VJ_{s}\cA^{(r,k)}$.

By applying an abstract result due to Kol\'a\v{r} (see
\cite{Kol83}) concerning a global decomposition formula for vertical 
morphisms, by using Proposition \ref{x} we can prove the 
following.

\bLm
Let $\lam$ be a Lagrangian and $\hat{\Xi}_V$ a variation vecrtor 
field. Let us set $\chi(\lam,\mathfrak{G}(\bar{\Xi})_{V})\byd
C^{1}_{1} (j_{2s}\hat{\Xi}\ten 
K_{hd\cL_{j_{2s}\bar{\Xi}_V}\lam})\equiv E_{j_{s}\hat{\Xi}\rfloor
hd\cL_{j_{2s+1}\bar{\Xi}_V}\lam}$. Let
$D_{H}$ be the horizontal differential on 
$J_{4s}\bY_{\zet}\ucar{\bX}VJ_{4s}\cA^{(r,k)}$.
Then we have:
\bEq\label{chi}
(\pi^{4s+1}_{2s+1})^{*}\chi(\lam,\mathfrak{G}(\bar{\Xi})_{V}) = 
E_{\chi(\lam,\mathfrak{G}(\bar{\Xi})_{V})} +
F_{\chi(\lam,\mathfrak{G}(\bar{\Xi})_{V})}\,,
\eEq
where
\beq
E_{\chi(\lam,\mathfrak{G}(\bar{\Xi})_{V}}: 
J_{4s}\bY_{\zet}\ucar{\bX}VJ_{4s}\cA^{(r,k)} \to
\Con^{*}_{0}[\cA^{(r,k)}]\ten\Con^{*}_{0}[\cA^{(r,k)}]\wed 
(\owed{n}T^{*}\bX) \,,
\eeq
and locally $F_{\chi(\lam,\mathfrak{G}(\bar{\Xi})_{V})} =
D_{H}M_{\chi(\lam,\mathfrak{G}(\bar{\Xi})_{V})}$ with
\beq
M_{\chi(\lam,\mathfrak{G}(\bar{\Xi})_{V})}:J_{4s}\bY_{\zet}\ucar{\bX}VJ_{4s}\cA^{(r,k)}\to 
\Con^{*}_{2s-1}[\cA^{(r,k)}]\ten \Con^{*}_{0}[\cA^{(r,k)}]\wed
(\owed{n-1}T^{*}\bX)\,.
\eeq
\eLm

\bPf
As a consequence of
linearity properties
of both $\chi(\lam, \Xi)$ and
  the Lie derivative operator $\pounds$ we have 
$\chi(\lam,\mathfrak{G}(\bar{\Xi})_{V}):
J_{2s}\bY_{\zet}\ucar{\bX}VJ_{2s}\cA^{(r,k)} \to
\Con^{*}_{2s}[\cA^{(r,k)}]\ten
\Con^{*}_{0}[\cA^{(r,k)}] \wed (\owed{n}T^{*}\bX)$ and
$D_{H}\chi(\lam,\mathfrak{G}(\bar{\Xi})_{V})$ $=$ $0$.
Thus Kol\'a\v r's decomposition formula can be applied. 
Coordinate expressions of both terms in the above invariant 
decomposition \eqref{chi} can be easily evaluated by a backwards 
procedure analogously to Example \ref{Ex1}.
\QED
\ePf

\bDf
We call the morphism $\cJ(\lam,\mathfrak{G}(\bar{\Xi})_{V})$ $\byd$
$E_{\chi(\lam,\mathfrak{G}(\bar{\Xi})_{V})}$  the {\em gauge-natural 
generalized Jacobi
morphism} associated with the Lagrangian $\lam$ and the variation 
vector field $\hat{\Xi}_{V}$.
\END\eDf

Notice that the morphism $\cJ(\lam,\mathfrak{G}(\bar{\Xi})_{V})$ is a
{\em linear} morphism with respect to the projection
$J_{4s}\bY_{\zet}\ucar{\bX}VJ_{4s}\cA^{(r,k)}\to J_{4s}\bY_{\zet}$.

For what we call the {\em gauge-natural second variational 
derivative} we have the following relations (see \cite{PaWi03}).

\bTh\label{comparison}
Let $\del^{2}_{\mathfrak{G}}\lam$ be the second  variational 
derivative of $\lam$ with respect to (prolongations of) vertical parts
of gauge-natural lifts of infinitesimal principal automorphisms. The following equalities
hold:
\bEq
\del^{2}_{\mathfrak{G}}\lam
= \mathfrak{G}(\bar{\Xi})_{V}\rfloor
\cE_{n}\left(\mathfrak{G}(\bar{\Xi})_{V}\rfloor\cE_{n}(\lam)\right)
=
\cE_{n}\left(\mathfrak{G}(\bar{\Xi})_{V}\rfloor
h(d\del\lam)\right)\,.
\eEq
\eTh

\section{Canonical covariant conserved currents}

In the following we assume that the field equations are generated by
means of a variational principle from a Lagrangian which is
gauge-natural invariant, \ie invariant with respect to any
gauge-natural lift of infinitesimal right invariant vector fields.

\bDf\label{gn}
Let $(\hat{\Xi},\xi)$ be a projectable vector field on $\bY_{\zet}$.
Let $\lam \in \Var^{n}_{s}$
be a generalized Lagrangian. We say $\hat{\Xi}$ to be a {\em symmetry\/}
of $\lam$ if $\cL_{j_{s+1}\hat{\Xi}}\,\lam = 0$.

We say $\lam$ to be a
{\em gauge-natural invariant Lagrangian} if the gauge-natural lift
$(\hat{\Xi},\xi)$ of {\em any} vector
field $\bar{\Xi} \in \cA^{(r,k)}$ is a  symmetry for
$\lam$, \ie if $\cL_{j_{s+1}\bar{\Xi}}\,\lam = 0$.
In this case the projectable vector field
$\hat{\Xi}\equiv \mathfrak{G}(\bar{\Xi})$ is
called a {\em gauge-natural symmetry} of $\lam$.\END
\eDf

The First Noether Theorem \cite{Noe18} takes a particularly interesting
form in the case of gauge-natural Lagrangians as shown by the following (see also the
fundamental classical reference \cite{Tra67}).

\bPr
\label{symmetry of L}
Let $\lam \in \Var^{n}_{s}$ be a gauge-natural Lagrangian and
$(\hat{\Xi},\xi)$
a gauge-natural symmetry of $\lam$. Then we have
$$
0 = - \pounds_{\bar{\Xi}} \rfloor \cE_{n}(\lam)
+d_{H}(-j_{s}\pounds_{\bar{\Xi}}
\rfloor p_{d_{V}\lam}+ \xi \rfloor \lam). $$
Suppose that
$(j_{2s+1}\sig)^{*}\left(- \pounds_{\bar{\Xi}} \rfloor \cE_{n}(\lam)\right) = 0$.
Then, the $(n-1)$--form
\beq
\eps = - j_{s}\pounds_{\bar{\Xi}} \rfloor p_{d_{V}\lam}+ \xi 
\rfloor \lam
\eeq
fulfills the equation $d \left((j_{2s}\sig)^{*}(\eps)\right) = 0$.
\ePr

If $\sig$ is a critical section for $\cE_{n}(\lam)$, \ie
$(j_{2s+1}\sig)^{*}\cE_{n}(\lam) = 0$, the above equation
admits a physical interpretation as a so-called {\em weak conservation law}
for the density associated with $\eps$.

\bDf
Let $\lam \in \Var^{n}_{s}$ be a gauge-natural Lagrangian and
$\bar{\Xi} \in \cA^{(r,k)}$. Then the sheaf morphism $\eps: 
J_{2s}\bY_{\zet} \ucar{\bX} VJ_{2s}\cA^{(r,k)} \to 
\cC_{2s}^{*}[\cA^{(r,k)}]\ten\,\cC_{0}^{*}[\cA^{(r,k)}]
\wed (\owed{n-1} T^{*}\bX)$
is said to be a {\em gauge-natural weakly conserved current\/}.\END
\eDf

\bRm\label{arbitrary1}
In general, this conserved current is not uniquely defined. In fact,
it depends on the choice of $p_{d_{V}\lam}$, which is not unique (see \eg
\cite{Vit98} and references quoted therein).
\END\eRm

\subsection{The deformed Lagrangian and Bianchi 
morphism}

In gauge-natural Lagrangian theories a well known procedure suggests
to perform suitable integrations by
parts to decompose the conserved current $\eps$ into the sum of
a conserved current vanishing along solutions of the Euler--Lagrange equations,
the so--called {\em reduced current} $\tilde{\eps}$, and the
formal divergence of a skew--symmetric (tensor) density called a {\em
superpotential} (which is defined modulo a divergence).
Within such a procedure, the generalized Bianchi identities
are in fact necessary and (locally) sufficient conditions for the 
conserved current
$\epsilon$ to be not only closed but also the divergence of a 
skew-symmetric (tensor) density
along solutions of the  Euler--Lagrange equations.

Let then $\lam$ 
be a gauge-natural invariant Lagrangian. We set
\bEq
\ome (\lam,\mathfrak{G}(\bar{\Xi})_{V}) \equiv  - \pounds_{\bar{\Xi}} 
\rfloor \cE_{n} (\lam): J_{2s}\bY_{\zet}
\to \Con_{2s}^{*}[\cA^{(r,k)}]\ten \, \Con_{0}^{*}[\cA^{(r,k)}]\wed (\owed{n}
T^{*}\bX) \,.
\eEq
The morphism $\ome (\lam,\mathfrak{G}(\bar{\Xi})_{V})$ so defined is 
a new `deformed' Lagrangian associated with the field equations of the
original Lagrangian $\lam$. It has been considered in applications 
\eg in General Relativity (see \cite{FFR03} and references quoted 
therein).
We have $D_{H}\ome(\lam,\mathfrak{G}(\bar{\Xi})_{V})= 0$ and by the 
linearity of the operator
$\pounds$ we can regard $\ome(\lam,\mathfrak{G}(\bar{\Xi})_{V})$ as 
the extended morphism
$\ome(\lam,\mathfrak{G}(\bar{\Xi})_{V}): J_{2s}\bY_{\zet} \ucar{\bX} 
VJ_{2s}\cA^{(r,k)}
\to \Con_{2s}^{*}[\cA^{(r,k)}]\ten\, \Con_{2s}^{*}[\cA^{(r,k)}]\ten\, 
\Con_{0}^{*}[\cA^{(r,k)}]\wed
(\owed{n}T^{*}\bX)$. 

The following Lemma (see \cite{PaWi03}) is a 
geometric version of the integration by parts procedure quoted above 
and it is based on a global decomposition formula of vertical 
morphisms due to Kol\'a\v{r} \cite{Kol83}.

\bLm\label{kol}
Let $\ome(\lam,\mathfrak{G}(\bar{\Xi})_{V})$ be as above.  On the 
domain of $\ome(\lam,\mathfrak{G}(\bar{\Xi})_{V})$ we have:
\beq
(\pi^{4s+1}_{s+1})^{*}\ome(\lam,\mathfrak{G}(\bar{\Xi})_{V}) = 
\bet(\lam,\mathfrak{G}(\bar{\Xi})_{V}) +
F_{\ome(\lam,\mathfrak{G}(\bar{\Xi})_{V})}\,,
\eeq
where
\bEq\label{bianchi}
&\bet(\lam,\mathfrak{G}(\bar{\Xi})_{V})\equiv
E_{\ome(\lam,\mathfrak{G}(\bar{\Xi})_{V})} : \\
& J_{4s}\bY_{\zet} \ucar{\bX} VJ_{4s}\cA^{(r,k)} \to
\Con_{2s}^{*}[\cA^{(r,k)}]\ten\,\Con_{0}^{*}[\cA^{(r,k)}]\ten\,\Con_{0}^{*}[\cA^{(r,k)}]\wed
(\owed{n}T^{*}\bX) \,
\eEq
and, {\em locally}, $F_{\ome(\lam,\mathfrak{G}(\bar{\Xi})_{V})} = 
D_{H}M_{\ome(\lam,\mathfrak{G}(\bar{\Xi})_{V})}$,
with
\beq
& M_{\ome(\lam,\mathfrak{G}(\bar{\Xi})_{V})}: \\
& J_{4s-1}\bY_{\zet} \ucar{\bX} VJ_{4s-1}\cA^{(r,k)}) \to
\Con_{2s}^{*}[\cA^{(r,k)}]\ten\,\Con_{2s-1}^{*}[\cA^{(r,k)}]\ten\,\Con_{0}^{*}[\cA^{(r,k)}]\wed 
(\owed{n-1}T^{*}\bX)\,.
\eeq
\eLm

\bDf
We call the global morphism $\bet(\lam,\mathfrak{G}(\bar{\Xi})_{V}) 
\byd E_{\ome(\lam,\mathfrak{G}(\bar{\Xi})_{V})}$
the {\em generalized} {\em Bianchi morphism} associated
with the Lagrangian $\lam$ and the variation vector field 
$\hat{\Xi}_{V}$.\END\eDf

\bRm
For any $(\bar{\Xi},\xi)\in \cA^{(r,k)}$, as a consequence of the 
gauge-natural invariance of the Lagrangian,
the morphism $\bet(\lam,\mathfrak{G}(\bar{\Xi})_{V}) \equiv
\cE_{n}(\ome(\lam,\mathfrak{G}(\bar{\Xi})_{V}))$ is {\em locally} 
identically vanishing.
We stress that these are just {\em local generalized Bianchi
identities} \cite{Ber58}.\END
\eRm

Let $\mathfrak{K}$
be the {\em kernel} of
  $\cJ(\lam,\mathfrak{G}(\bar{\Xi})_{V})$.
As a consequence of Theorem \ref{comparison} and the considerations
above we have the following result, the detailed proof of which will 
appear in \cite{PaWi03}.

\bTh\label{giacomino}
The generalized Bianchi morphism is globally vanishing for the 
variation vector field $\hat{\Xi}_{V}$ if and only
if $\del^{2}_{\mathfrak{G}}\lam\equiv\cJ(\lam,\mathfrak{G}(\bar{\Xi})_{V})=
0$, \ie if and only if
$\mathfrak{G}(\bar{\Xi})_{V}\in\mathfrak{K}$.
\eTh


\subsection{A strong conservation law equivalent to the Jacobi 
equations}

 From now on we shall write $\ome(\lam,
\mathfrak{K})$ to denote $\ome(\lam,\mathfrak{G}(\bar{\Xi})_{V})$ 
when $\mathfrak{G}(\bar{\Xi})_{V}$ belongs to $\mathfrak{K}$. 
Analogously
for $\bet$ and other morphisms.

First of all let us make the following important considerations 
(partial results in this direction were already given in 
\cite{FPW03}, where the consequences on the concept of curvature of a 
gauge-natural invariant principle have been stressed).
\bPr
 Let
$D_{V}$ be the vertical differential on 
$J_{4s}\bY_{\zet}\ucar{\bX}VJ_{4s}\cA^{(r,k)}$.
For each $\bar{\Xi}\in \cA^{(r,k)}$ such that $\bar{\Xi}_{V}\in 
\mathfrak{K}$, we have
\bEq
\cL_{j_{s}\bar{\Xi}_{H}}\ome(\lam,
\mathfrak{K})=-D_{H}(-j_{s}\pounds_{\bar{\Xi}_{V}}
\rfloor p_{D_{V}\ome(\lam,\mathfrak{K})}) \,.
\eEq
\ePr

\bPf
The horizontal splitting gives us 
$\cL_{j_{s}\bar{\Xi}}\ome(\lam,\mathfrak{K})=\cL_{j_{s}\bar{\Xi}_{H}}\ome(\lam,\mathfrak{K}) 
+
\cL_{j_{s}\bar{\Xi}_{V}}\ome(\lam,\mathfrak{K})$. Furthermore, 
$\ome(\lam,\mathfrak{K})\equiv - \pounds_{\bar{\Xi}} \rfloor
\cE_{n}(\lam) = \cL_{j_{s}\bar{\Xi}}\lam - d_{H}(-j_{s}\pounds_{\bar{\Xi}}
\rfloor p_{d_{V}\lam}+ \xi \rfloor \lam)$; so that
\beq
\cL_{j_{s}\bar{\Xi}_{V}}\ome(\lam,\mathfrak{K})=\cL_{j_{s}\bar{\Xi}_{V}}\cL_{j_{s}\bar{\Xi}}\lam=\cL_{j_{s}[\bar{\Xi}_{V},
\bar{\Xi}_{H}]}\lam\,.
\eeq
  On the other hand we have
\beq
\cL_{j_{s}\bar{\Xi}_{H}}\ome(\lam,\mathfrak{K}) = \cL_{j_{s}[\bar{\Xi}_{H},
\bar{\Xi}_{V}]}\lam=-\cL_{j_{s}\bar{\Xi}_{V}}\ome(\lam,\mathfrak{K})\,.
\eeq

Recall now that from the Theorem above we have $\bar{\Xi}_{V}\in
\mathfrak{K}$ if and only if $\bet(\lam,\mathfrak{K})=0$.
Since
\beq
& &\cL_{j_{s}\bar{\Xi}_{V}}\ome(\lam,\mathfrak{K})=
  - \pounds_{\bar{\Xi}_{V}} \rfloor \cE_{n}(\ome(\lam,\mathfrak{K}))
+D_{H}(-j_{s}\pounds_{\bar{\Xi}_{V}}
\rfloor p_{D_{V}\ome(\lam,\mathfrak{K})}) = \\
& &   = \bet(\lam,\mathfrak{K})+D_{H}(-j_{s}\pounds_{\bar{\Xi}_{V}}
\rfloor p_{D_{V}\ome(\lam,\mathfrak{K})})\,,
\eeq
  we get the assertion.
\END
\ePf

It is easy to realize that, because of the gauge-natural invariance 
of the generalized
Lagrangian $\lam$, the new generalized Lagrangian 
$\ome(\lam,\mathfrak{K})$ is gauge-natural invariant too, \ie
$\cL_{j_{s}\bar{\Xi}}\,\ome(\lam,
\mathfrak{K})=0$.

Even more, we can state the following:
\bPr
Let $\bar{\Xi}_{V}\in
\mathfrak{K}$. We have
\bEq
\cL_{j_{s}\bar{\Xi}_{H}}\ome(\lam,
\mathfrak{K})=0\,.
\eEq
\ePr

\bPf
In fact, when $\bar{\Xi}_{V}\in
\mathfrak{K}$, since $\cL_{j_{s}\bar{\Xi}_V}\ome(\lam,
\mathfrak{K})=[\bar{\Xi}_V,\bar{\Xi}_V]\rfloor \cE_n (\lam)+ 
\bar{\Xi}_V \rfloor\cL_{j_{s}\bar{\Xi}_V}\cE_n (\lam) =0$, we have
\beq
& &0=\cL_{j_{s}\bar{\Xi}}\ome(\lam,
\mathfrak{K})= \cL_{j_{s}\bar{\Xi}_V}\ome(\lam,
\mathfrak{K})+\cL_{j_{s}\bar{\Xi}_H}\ome(\lam,
\mathfrak{K})= \cL_{j_{s}\bar{\Xi}_H}\ome(\lam,
\mathfrak{K})\,.
\eeq
\QED
\ePf

As a quite relevant byproduct we get also the following (this result 
can be compared with \cite{CFT96}, where some preliminary results have 
been obtained for second order Lagrangians; compare also with partially analogous results
given in \cite{FFR03}).

\bCr
Let $\bar{\Xi}_{V}\in
\mathfrak{K}$. We have the {\em covariant} strong conservation law
\bEq
D_{H}(-j_{s}\pounds_{\bar{\Xi}_{V}}
\rfloor p_{D_{V}\ome(\lam,\mathfrak{K})})=0\,.
\eEq
\eCr

\bPf
\beq
  0 &=&\cL_{j_{s}\bar{\Xi}_H}\ome(\lam,
\mathfrak{K})= -\bet(\lam,\mathfrak{K}) -D_{H}(-j_{s}\pounds_{\bar{\Xi}_{V}}
\rfloor p_{D_{V}\ome(\lam,\mathfrak{K})}) = \\
&=& - D_{H}(-j_{s}\pounds_{\bar{\Xi}_{V}}
\rfloor p_{D_{V}\ome(\lam,\mathfrak{K})})\,.
\eeq
\QED\ePf

\bDf
We define the covariant $n$-form
\bEq
\cH(\lam,\mathfrak{K})=-j_{s}\pounds \,\rfloor \, 
p_{D_{V}\ome(\lam,\mathfrak{K})}
\,,
\eEq
  to be a Hamiltonian form for
$\ome(\lam,\mathfrak{K})$ on the Legendre bundle $\Pi\equiv 
V^{*}(J_{2s}\bY_{\zet} \ucar{\bX} VJ_{2s}\cA^{(r,k)})\wed (\owed{n-1}
T^{*}\bX)$ (see \cite{MaSa00}).
\END
\eDf

Let $\Ome$ be the multisimplectic form on $\Pi$.
It is well known that every Hamiltonian form $\cH$ admits a 
Hamiltonian connection $\gam_{\cH}(\lam,\mathfrak{K})$ such that
$
\gam_{\cH}(\lam,\mathfrak{K})\rfloor\Ome = d\cH(\lam,\mathfrak{K})$. 
Let then  $\gam_{\cH}(\lam,\mathfrak{K})$ be the corresponding 
Hamiltonian connection form (see \cite{MaSa00}).

As it has been stressed in \cite{FrPa01} the Euler--Lagrange 
equations together with the Jacobi
equations for a given Lagrangian $\lam$ can be obtained out of a 
unique variational
problem for
the Lagrangian $\del\lam$. This can be performed by requiring the 
invariance of $\del\lam$ with respect to vertical parts of 
gauge-natural lifts of infinitesimal principal  automorphisms, which 
are solutions of the classical Jacobi equations
along critical sections, thus recovering a well known classical 
result (see \eg \cite{Vari,CFT96,GoSt73,Tau69}).

We can finally formulate and prove the following important new result:

\bTh 
For all $\mathfrak{G}(\Xi)_V \in \mathfrak{K}$ the Hamilton
equations for the
Hamiltonian connection form $\gam_{\cH}(\lam,\mathfrak{K})$ coincide 
with the kernel of the generalized gauge-natural Jacobi 
morphism.
\eTh

\bPf
The Hamilton
equations for the Hamiltonian connection 
$\gam_{\cH}(\lam,\mathfrak{K})$ are identically satisfied being 
equivalent to the kernel of the Euler lagrange morphism of  the 
Lagrangian $\ome(\lam,\mathfrak{K})$ (see \eg \cite{MaSa00}), which 
coincides with $\mathfrak{K}$ because of Theorem \ref{giacomino}. 
Generalized Bianchi identities appear then as constraints for such an 
equivalence to hold true (see also 
\cite{GoSt73}).
\QED
\ePf

\bEx
Let $Q$ be a $n$--dimensional manifold and $(TQ,Q,\tau_Q)$ its tangent
bundle. It is well known that $TQ$ is a natural bundle associated to 
the principal bundle $L(\R)$ of frames in $\R$; the gauge-natural 
lift order is $(r,k)= (0,1)$.  In this case the natural lift of 
tangent vector fields to $\R$ is the tangent lift and will be denoted 
by a dot. Let $g$ be a Riemannian metric on $Q$. The {\em geodesics} 
of $(Q,g)$
are those curves $\gam : \R \to Q$ whose tangent vector $\dot \gam$ is parallel
along $\gam$, \ie it satisfies $\nab_{\dot{\gam}}\dot{\gam}=0$.
We assume that the {\em Jacobi fields} of $(Q,g)$ are those 
vectorfields $\eta = \eta^i \der_q^i$, with $\eta^i = \xi^i-\dot{q}^i 
\xi^0$, \ie vertical parts of tangent vector fields $\xi = \xi^0 
\der_t +\xi^i\der_{q^{i}}$
  and characterized along geodesics $\gam$ by the differential equation:
\beq
\nab^2_{\dot \gam} \eta +Riem_g(\eta,\dot \gam ,\dot \gam ) = 0 \, ,
\eeq
  where $\nab^2_{\dot\gam}$ denotes the second--order covariant derivative
along the curve $\gam$ and $Riem_g$ is the tri--linear mapping defining the
Riemannian curvature of $g$. Jacobi fields generically define infinitesimal
deformations of geodesics into families of nearby geodesics. The metric $g$ can be lifted to
a metric $g^C$ on the  manifold $TQ$,
called the ``complete lift'', as follows.

 Let $g=g_{ij}dq^i dq^j$ in a local
chart $(U,q^i)$; then the corresponding local expression of $g^C$ in
$(TU;q^i,u^i)$ is
$ g^C = 2 g_{ij} \del u^i dq^j$, where $\del u^i$ stands for $ \del 
u^i = du^i + \Gam^i_{mk} u^m dq^k$.

For any function $f: Q \to \R$ a new function $\der f: TQ\to\R$ is defined by
setting locally: $(\der f)(q^i,u^i) \equiv (\der_j f) u^j $.
With this notation $g^C$ can be locally expressed by:
$ g^C = (\der g_{ij}) du^i du^j +2 g_{ij} du^i dq^j $.
We can easyly 
see that the
system formed by the geodesic equation of $g$ in $Q$ and the Jacobi equation
associated with $g$ in $TQ$ is the geodesic equation in $TQ$ of the complete
lift metric $g^C$. Therefore this system follows from a variational principle
on $TQ$ based on the energy functional defined by the lifted metric $g^C$.
The energy functional of $g$ is based on the
Lagrangian $\lam = \frac{1}{2} g_{ij}u^i u^j $;
the  associated first--order deformed Lagrangian is thence given by 
$\ome = g_{ij}[ \dot{\eta}^i + \Gam^i_{mk} u^m \eta^k ] u^j$.

Then $\ome$ is in fact the energy Lagrangian of the lifted metric
$g^C=2g_{ij} \del u^i dq^j$. Notice that similar results have been 
obtained in \cite{Vari,CFT96};  however, we stress that here Jacobi 
fields, on which $\ome$ depends, are assumed to be of a specific 
nature, \ie vertical parts of tangent lifts and they have to satisfy 
identically the Jacobi equation associated with $g^C$ in $TTQ$ 
defining generalized Bianchi identities for $Riem_{g^C}$.\END

\eEx

\subsection{Concluding 
remarks}

It is interesting and useful to compare our results  with those of the recent
interesting paper 
\cite{GMS03}, where conservation laws associated with generalized 
Lagrangian symmetries, and in particular with divergence symmetries 
of a Lagrangian, are presented in the framework of infinite order 
variational bicomplexes (for finite order variational sequences 
partial results in the same direction have been already obtained in 
\cite{Kru93}). 

We notice in particular that variation vector fields 
$\mathfrak{G}(\bar{\Xi})_V$ which we are considering here are in fact 
generalized symmetries of $\cE(\lam)$ in the sense of \cite{cacca} and
\cite{GMS03}.
Moreover, let us once more recall that one can represent second 
variational derivatives as iterated variational Lie derivatives of the 
Lagrangian and in particular relate them to the Lie derivative of 
Euler Lagrange morphisms (see also 
\cite{FrPa00,FrPa01,FPV05,PaWi03,PaWi04}).
  As a consequence of 
Theorem \ref{comparison}, we can now  provide an intrinsic interpretation of 
the last term appearing in the {\em characteristic equation} for 
generalized symmetries derived in \cite{GMS03}. From Eq. $(18)$ of 
\cite{FPV05}, in fact, it is easy to realize that such term is nothing but the 
difference between the second variational derivative of $\lam$ and 
the Jacobi morphism (the latter being precisely characterized as the 
vertical differential of $\cE_n (\lam)$), {\em up to divergences} - 
and it vanishes along critical sections.

 In other words, the 
characteristic equation in our case can be written, up to 
divergences, as 
follows:
\beq
\del_\mathfrak{G}\cE(\lam)-\cE(\del_\mathfrak{G}\lam)=[\del^2_\mathfrak{G}\lam 
-\cJ(\lam,\mathfrak{G}(\bar{\Xi})_{V}) ]   \equiv 0\, .
\eeq
where 
$[\,]$ denotes the equivalence class in the variational sequence. In 
the finite order variational sequence such an equivalence class is 
vanishing by definition also along non critical sections, being the 
equivalence class of the horizontal differential of contact forms of 
higher degree (this fact has been already  stressed in \cite{PaWi03}).

We also 
notice that by adapting the results of  \cite{GMS03} to our case we 
would have that {\em requiring} that  $\cE(\ome(\lam, 
\mathfrak{G}(\bar{\Xi})_V)) = 0$, then $\mathfrak{G}(\bar{\Xi})_V$ 
should be a divergence symmetry of $\lam$, \ie 
$\cL_{\mathfrak{G}(\bar{\Xi})_V}\lam$ should be a total divergence. 
We remark  that - because of the gauge-natural invariance of $\lam$ 
- one finds $\cL_{\mathfrak{G}(\bar{\Xi})_V}\cE(\lam)$ $=$ 
$\cL_{\mathfrak{G}(\bar{\Xi})}\cE(\lam)=0$. Since - by Theorem 
\ref{giacomino} - $\A \,\, \mathfrak{G}(\Xi)_V$ $\in$ $\mathfrak{K}$, 
$\cE(\ome(\lam, \mathfrak{K}))=0$, we can conclude that 
$\cL_{\mathfrak{G}(\bar{\Xi})_V}\lam$ is always a total divergence on 
the kernel of the generalized gauge-natural Jacobi 
morphism, so providing new important insights on the results of \cite{GMS03}.

\subsection{Acknowledgements}
Thanks are due to I. 
Kol\'a\v{r}  for useful discussions. Special thanks are due to R. 
Vitolo, collaborator and member of a common research program on 
higher variations, for helpful remarks.
The second author (M.P.) 
especially thanks I. Kol\'a\v{r} for the kind invitation to take part 
to the Conference. 

The leading idea of the paper has took its first 
form during the stay of M.P. at the Mathematical Institute of the 
Silesian University in Opava (Czech Republic), May-October 2000, 
supported by University of Torino and the Italian Council of Researches through grant n. 
203.01.71/03.01.02 and under the supervision of D. Krupka; the 
scientific contacts had with him, although not directly connected 
with the second variational derivative, are therefore here gratefully 
acknowledged. 


\end{document}